\documentclass[pdflatex,sn-mathphys-ay]{sn-jnl}

\usepackage{graphicx}%
\usepackage{multirow}%
\usepackage{amsmath,amssymb,amsfonts}%
\usepackage{amsthm}%
\usepackage{mathrsfs}%
\usepackage[title]{appendix}%
\usepackage{xcolor}%
\usepackage{textcomp}%
\usepackage{manyfoot}%
\usepackage{booktabs}%
\usepackage{algorithm}%
\usepackage{algorithmicx}%
\usepackage{algpseudocode}%
\usepackage{listings}%

\usepackage{subfigure}

\usepackage{comment}

\theoremstyle{thmstyleone}%
\theoremstyle{thmstyletwo}%

\theoremstyle{thmstylethree}%

\begin{document}

\newcommand{\flagcomment}[1]{\textcolor{green}{#1}}

\title[Revisiting Information Diffusion Beyond Explicit Social Ties]{Revisiting Information Diffusion Beyond Explicit Social Ties: A Study of Implicit‑Link Diffusion on Twitter}

\author*[1]{\fnm{Yuto} \sur{Tamura}}\email{yuto.tamura@snlab.cs.tsukuba.ac.jp}
\author[2]{\fnm{Sho} \sur{Tsugawa}}\email{s-tugawa@cs.tsukuba.ac.jp}
\author[3]{\fnm{Kohei} \sur{Watabe}}\email{kwatabe@mail.saitama-u.ac.jp}

\affil*[1]{%
  \orgdiv{Graduate School of Science and Technology}, 
  \orgname{University of Tsukuba}, 
  \orgaddress{\city{Tsukuba}, \postcode{305--8573}, \country{Japan}}%
}

\affil[2]{%
  \orgdiv{Institute of Systems and Information Engineering},
  \orgname{University of Tsukuba}, 
  \orgaddress{\city{Tsukuba}, \postcode{305--8573}, \country{Japan}}%
}

\affil[3]{%
  \orgdiv{Graduate School of Science and Engineering}, 
  \orgname{Saitama University}, 
  \orgaddress{\city{Saitama}, \postcode{338--8570}, \country{Japan}}%
}

\abstract{
Information diffusion on social media platforms is often assumed to occur primarily through explicit social connections, such as follower or friend ties. However, information frequently propagates beyond these observable ties—through external websites, search engines, or algorithmic recommendations—creating implicit links. How the presence of implicit links affects the diffusion process remains unclear.
In this study, we investigate the characteristics of implicit links on Twitter using four large-scale datasets. Our analysis reveals that users who are farther from the original source in the social network are more likely to engage in diffusion via implicit links. Although implicit links contribute less to the overall diffusion volume than explicit links, they play a distinct role in disseminating content across diverse and topologically distant communities.
We further examine the user attributes associated with the formation of implicit links and show that these features are unevenly distributed across the network and exhibit moderate levels of homophily and monophily. 
Together, these findings demonstrate that implicit links exert a meaningful influence on information diffusion and highlight the importance of incorporating them into models of diffusion and social influence.}




\maketitle

\section{Introduction}
\label{section:introdcution}


Analyzing and understanding the characteristics of information diffusion on social media has been an important research topic, enabling various applications such as limiting the spread of misinformation and optimizing viral marketing strategies.
Information posted by companies or individuals on social media has the potential to spread extensively. Consequently, social media is utilized as a platform for implementing viral marketing strategies that leverage word-of-mouth~\citep{marketing_through_instagram_influencers}. In contrast, the ease of information dissemination inherent in social media also leads to the proliferation of fake news, posing significant societal challenges~\citep{the_covid_19_infodemic_twitter_versus_facebook}. Therefore, there is a growing body of research aimed at 
identifying the characteristics of information diffusion and designing interventions to either promote or suppress its spread.
For instance, prior studies have proposed methods to identify influential users~\citep{temporal_sequence_of_retweets_help_to_detect_influential_nodes_in_social_networks,a_sharpley_value_based_approach_to_discover_influential_nodes_in_social_networks}, or to control the size of diffusion by adding or removing network links~\citep{Empirical_evaluation_of_link_deletion_methods_for_limiting_information_diffusion_on_social_media,gelling_and_melting_large_graphs_by_edge_manipulation}.

\begin{figure}[tbp]
\centering
\subfigure[Follow Network] {
    \includegraphics[width=0.35\textwidth]{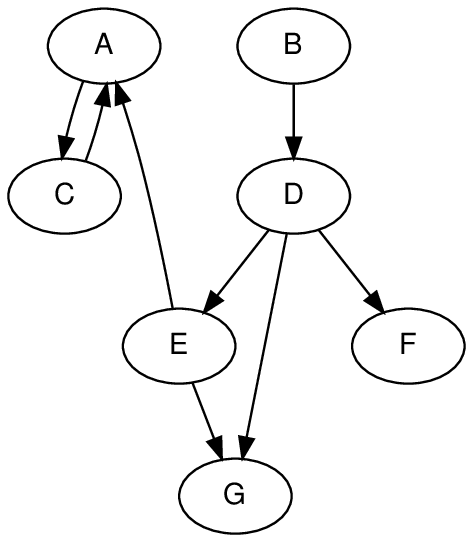}
    \label{figure:following_graph}
  }
\subfigure[Cascade Graph] {
\includegraphics[width=0.35\textwidth]{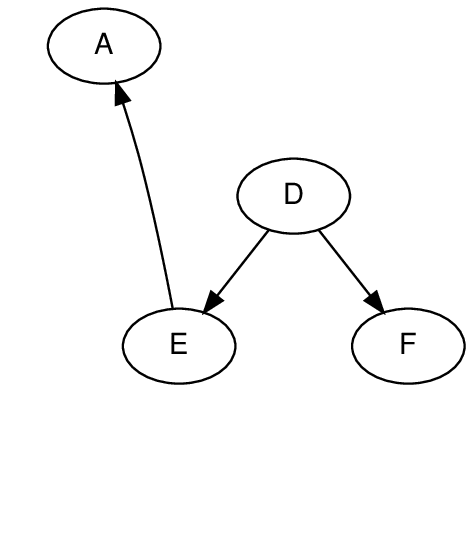}
  \label{figure:cascade_graph}
}
\caption{
Example of a follow network and cascade graph:
(a) An illustration of a follow network where each directed link represents a follow relationship. Here, $B \to D$ indicates that $D$ is a follower of $B$.
(b) An example of a cascade graph when a post by node D is reposted by nodes A, E, and F. Given that nodes E and F follow node D (as depicted in Fig.~\ref{figure:following_graph}), the post is considered to be disseminated from node D to nodes E and F, thus resulting in the cascade graph having links $(D, E)$ and $(D, F)$. Similarly, since node A follows node E, the post is considered to be disseminated from node E to A, and the cascade graph has a link $(E, A)$.
}
    \label{figure:following_and_cascade}
\end{figure}

Most existing studies assume that information is transmitted through explicit relationships (i.e., follow relationships) between users. Under this assumption, diffusion is typically represented using a cascade graph, 
where nodes represent users and a directed link $(u, v)$ indicates that information was disseminated from user $u$ to user $v$~\citep{information_diffusion_in_online_social_networks_a_survey}. On platforms like Twitter, if user $v$ follows user $u$ and reposts user $u$'s content, it is natural to infer that the information was passed from $u$ to $v$. Cascade graphs are constructed based on such assumptions.
Figure~\ref{figure:following_and_cascade} shows an example of a follow network and a corresponding cascade graph illustrating information diffusion over the network.

\begin{figure}[tbp]
\centering
\subfigure[Follow Network] {
    \includegraphics[width=0.35\textwidth]{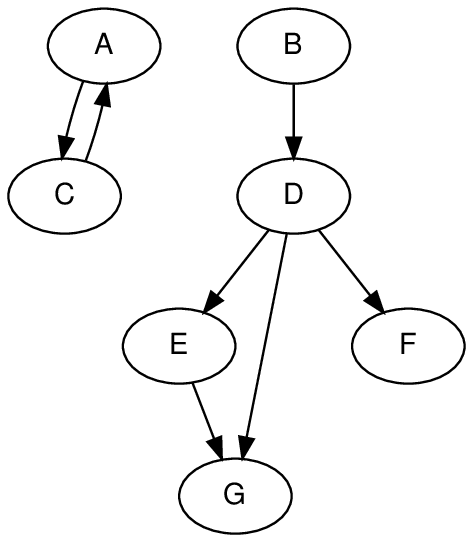}
    \label{figure:imp_following_graph}
  }
\subfigure[Cascade Graph] {
\includegraphics[width=0.35\textwidth]{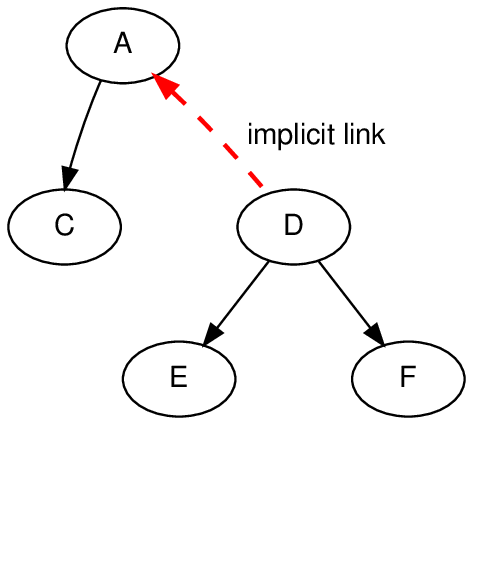}
  \label{figure:imp_cascade_graph}
}
\caption{
Example of a follow network and a cascade graph demonstrating nontrivial information diffusion via an implicit link:
(a) An illustration of a follow network where each directed link represents a follow relationship. Here, $A \to C$ indicates that $C$ is a follower of $A$.
(b) An example of a cascade graph where a post by node D is reposted by nodes A, C, E, and F. While E and F follow D, A does not. The pathway by which the information reached A is not traceable through explicit links and is thus considered to be implicit. Since C follows A, it can be inferred that C received the information from A through an explicit link.
}
\label{figure:imp_following_and_cascade}
\end{figure}

However, information diffusion does not always occur solely through explicit follow relationships. Users may encounter posts through external sources such as websites, search engines, or trending topics, and spontaneously repost them even without direct social ties~\citep{luceri202IAR_SAR}. In such cases, the diffusion paths cannot be reconstructed solely from the follow network, resulting in disconnected cascade graphs. Figure~\ref{figure:imp_following_and_cascade} shows an example. Here, node A reposts content from node D despite the absence of any follow path, making it unclear how A discovered the post. We hypothesize that such nontrivial diffusion occurs via \emph{implicit links}, which reflect information paths not captured by observable social ties.
Prior research has shown that diffusion cascades constructed from repost histories and follow networks often result in disconnected graphs~\citep{Information_Diffusion_and_External_Influence_in_Networks,Inferring_Missing_Retweets_in_Twitter_Information_Cascades,Existence_of_Twitter_Users_with_Untraceable_Retweet_Paths_and_its_Implications,Information_Diffusion_and_Provenance_of_Interactions_in_Twitter_Is_It_Only_about_Retweets_}. This observation suggests that real-world diffusion frequently involves implicit links that cannot be directly inferred from follow relationships.

Despite the prevalence of such nontrivial paths, the characteristics of implicit link diffusion have received limited attention. Understanding how implicit links function is essential for constructing more realistic diffusion models. Existing models predominantly assume that information propagates through observable social ties, thereby neglecting nontrivial pathways. Incorporating the dynamics of implicit link diffusion could significantly enhance the accuracy and applicability of such models.

In this paper, we revisit the dynamics of information diffusion on social media, focusing specifically on nontrivial diffusion via implicit links. Using large-scale Twitter datasets of repost cascades, we formulate the following research questions (RQs) and explain the motivations underlying each:

\begin{itemize}
	\item {\bf (RQ1)}: How does nontrivial information diffusion via implicit links occur on social media?\\Understanding these mechanisms is essential for building realistic diffusion models and separating platform-mediated from exogenous exposure.
	\item {\bf (RQ2)}: To what extent do reposts via implicit links affect the size and structure of diffusion cascades?\\Quantifying their effect on cascade size is essential for accurately assessing the impact of implicit links on diffusion.
	\item {\bf (RQ3)}: To what extent do implicit links facilitate information diffusion across community boundaries?\\
    Examining whether information confinement within communities—a phenomenon typical of social media—also appears in implicit-link diffusion is essential for evaluating the qualitative effects of implicit links on cascades.
	\item {\bf (RQ4)}: What are the behavioral and structural characteristics of users who engage in or induce implicit link diffusion?\\
    Identifying users’ susceptibility and inducement contributes to a micro-level understanding of how implicit-link diffusion operates and influences information spread.

\end{itemize}

To answer these questions, we conduct a comprehensive analysis across multiple datasets of repost cascades on Twitter~\citep{temporal_sequence_of_retweets_help_to_detect_influential_nodes_in_social_networks,the_anatomy_of_a_scientific_rumor,emprical_analysis_of_the_relation_between_comunity, Turkish_ds}.
These datasets were previously collected via the Twitter API, are publicly available, and contain both follower ties (i.e., who follows whom) and repost relations (i.e., who reposted whose post and when).

We analyze the relationship between the source and reposting users to assess diffusion distance ({\bf RQ1}), evaluate the impact of reposts on cascade growth ({\bf RQ2}), and investigate the extent to which reposts via implicit links facilitate inter-community diffusion ({\bf RQ3}). Additionally, we examine user-level behavioral patterns to understand who is more susceptible to or likely to induce implicit link diffusion ({\bf RQ4}).

Across four Twitter datasets, we find that (i) the probability that a repost occurs via an implicit link increases with geodesic distance from the source; (ii) explicit-link reposts expand the cascade more than implicit-link reposts by generating a larger number of downstream reposts; (iii) implicit-link reposts cross community boundaries more frequently than explicit-link reposts; and (iv) user-level susceptibility to tweets from their followees, susceptibility to tweets that have not been adopted by their followees, and the degree to which their own posts are reposted via implicit links, all exhibit weak-to-moderate homophily together with monophily.

This paper is an extended version of our previous conference paper~\citep{asonam_proc}, which provided preliminary analyses of nontrivial diffusion. In this extended version, we expand the analysis in two key directions:

\begin{itemize}
 \item     {\bf Broader Dataset Coverage}: In addition to the three datasets analyzed in our conference version—Higgs, Nepal, and Ordinary—we include a newly added Turkish dataset to assess the generalizability of our findings across different social events and geography.
\item    {\bf User-Level Behavioral Analysis}: We introduce new analyses to examine individual-level susceptibility and influence in implicit link diffusion. Leveraging metrics such as the \textit{Spontaneous Adoption Rate} (SAR) and \textit{Influence-driven Adoption Rate} (IAR)~\cite{luceri202IAR_SAR}, we investigate the user-level susceptibility to implicit link and explicit link diffusions.
We also introduce a new metric, \textit{Repost via Explicit link Rate} (RER), to explore users who induce implicit link diffusion.

\end{itemize}

The remainder of this paper is organized as follows:  
Section~\ref{section:related_work} reviews related work.  
Section~\ref{section:perliminarie} introduces key definitions and describes the datasets.  
Section~\ref{section:analysis_and_discussion} presents our analysis of implicit link diffusion ({\bf RQ1-RQ3}).   
Section~\ref{section:user_sucepibility_and_influence} examines user susceptibility and influence ({\bf RQ4}).  
Section~\ref{sec:discuss} discusses broader implications, and  
Section~\ref{section:conclusion} concludes the paper.

\section{Related Work}
\label{section:related_work}

Researchers have extensively investigated the dynamics of information diffusion on social media platforms~\citep{temporal_sequence_of_retweets_help_to_detect_influential_nodes_in_social_networks,the_anatomy_of_a_scientific_rumor,Why_Rumors_Spread_so_Quickly_in_Social_Networks,twitter_adoption_and_use_in_mass_convergence_and_emergency_events,a_sharpley_value_based_approach_to_discover_influential_nodes_in_social_networks,understanding_communication_dynamics_on_twitter_during_natural_disasters,Negative_Messages_Spread_Rapidly_and_Widely_on_Social_Media,The_spread_of_true_and_false_news_online}. For instance, \cite{The_spread_of_true_and_false_news_online} found that false information tends to spread more widely and rapidly than true information. Similarly, \cite{Negative_Messages_Spread_Rapidly_and_Widely_on_Social_Media} demonstrated that negative messages propagate faster and farther than neutral or positive ones.  In a similar vein, \cite{Mathew_spread_of_hatespeach} and \cite{Masud_hate_topic_repost_prediction} analyzed the diffusion of hate speech on social media.
Researchers have also analyzed diffusion patterns in specific contexts. For example, \cite{the_anatomy_of_a_scientific_rumor} investigated how the announcement of the Higgs boson discovery spread on Twitter. Other studies have examined information dissemination during elections~\citep{twitter_adoption_and_use_in_mass_convergence_and_emergency_events}, natural disasters~\citep{understanding_communication_dynamics_on_twitter_during_natural_disasters}, and global health crises such as the COVID-19 pandemic~\citep{the_covid_19_infodemic_twitter_versus_facebook}.

Insights into information diffusion mechanisms have led to various practical applications, such as influencer identification and strategies to control the scale of information spread. A large body of work has focused on identifying influential users by leveraging either the network topology~\citep{a_sharpley_value_based_approach_to_discover_influential_nodes_in_social_networks} or diffusion history~\citep{temporal_sequence_of_retweets_help_to_detect_influential_nodes_in_social_networks}. Additionally, \cite{gelling_and_melting_large_graphs_by_edge_manipulation} proposed network modification techniques to suppress diffusion, while \cite{Empirical_evaluation_of_link_deletion_methods_for_limiting_information_diffusion_on_social_media} empirically showed that link deletion strategies are often ineffective in real-world diffusion cascades. They argued that one reason for this ineffectiveness is the presence of nontrivial diffusion paths outside explicit follow‑network ties.

While relatively few studies have addressed these nontrivial diffusion paths, a number of studies have investigated related phenomena from different perspectives. \cite{Inferring_Missing_Retweets_in_Twitter_Information_Cascades} noted that data collection limitations and privacy settings can hinder the accurate reconstruction of diffusion paths. \cite{Information_Diffusion_and_Provenance_of_Interactions_in_Twitter_Is_It_Only_about_Retweets_} found that approximately 50\% of reposts are made by users with explicit ties, with an additional 13\% explained by other interactions such as quotes. \cite{Information_Diffusion_and_External_Influence_in_Networks} reported that nearly 29\% of diffusion is influenced by sources external to the follow network. \cite{Existence_of_Twitter_Users_with_Untraceable_Retweet_Paths_and_its_Implications} introduced the concept of untraceable users, who repost content via paths that cannot be inferred from observable social ties. They showed a negative correlation between the number of such users and the popularity (follower count) of the original poster.
While most of these studies focus on the frequency of such nontrivial diffusion paths, our work goes further by analyzing how implicit link diffusion affects overall diffusion dynamics and structure, particularly in contrast with diffusion via explicit links.

The most closely related work to our study is by \cite{luceri202IAR_SAR}, who proposed a framework to quantify user susceptibility to diffusion via implicit links. They defined reposts made without explicit social ties as spontaneous adoptions and those made via known social ties as influence-driven adoptions. Their analysis showed that users who are more susceptible to either type of adoption tend to be adjacent to similarly susceptible users, suggesting a homophilous structure.
Building on this framework, our study analyzes user susceptibility and homophily using different datasets. While \cite{luceri202IAR_SAR} approximated explicit relationships using mention interactions, we use actual follow relationships as a direct representation of explicit social ties. This complements Luceri et al.’s claims by examining a different type of social connection.
In addition to analyzing user susceptibility, we also investigate users who are more likely to induce implicit link diffusion using a newly proposed metric (Section~\ref{section:user_influence}), and we analyze the structural and dynamic effects of implicit links on diffusion patterns (Section~\ref{section:analysis_and_discussion}). Taken together, our study provides a complementary and expanded understanding of implicit-link diffusion, both at the user level and at the cascade level.

\section{Preliminaries}
\label{section:perliminarie}

\subsection{Terminologies and Notations}

We define a follow network as a directed graph $G = (V, E)$, where $V$ denotes the set of nodes representing social media users, and $E$ denotes the set of directed links representing follow relationships. The set of users followed by user $u$ is denoted as $\Gamma(u)$. For a given post $t$, its author is denoted by $a(t)$, and a repost of $t$ by user $v$ is denoted as $r(t, v)$. Each original post and its associated reposts constitute a \textit{diffusion cascade}.

We represent the structure of a diffusion cascade using a diffusion graph $H_t = (R_t, E_t)$, where $R_t$ is the set of users who either posted or reposted $t$, and $E_t$ is the set of directed links representing inferred diffusion paths. An edge $(u, v) \in E_t$ exists if user $v$ follows user $u$ (i.e., $u \in \Gamma(v)$) and reposts $t$ after $u$ posted or reposted it—i.e., the timestamp of $r(t, v)$ is later than that of $r(t, u)$ or $a(t) = u$.

To distinguish between diffusion occurring through social ties and diffusion without such ties, we introduce two key concepts: \textit{explicit links} and \textit{implicit links}. A repost $r(t, u)$ is considered to be propagated via an explicit link if there exists a preceding node $v$ such that $(v, u) \in E_t$ in the diffusion graph—i.e., if the repost path is traceable through the follow network. Conversely, if no such $v$ exists (i.e., $u$ has no incoming edges in $H_t$), then $r(t, u)$ is considered to be propagated via an implicit link, implying that the diffusion path cannot be explained by observable follow relationships. That is, this classification uses only the presence or absence of a 1-hop predecessor of user $u$ in $H_t$; the existence of any longer (multi-hop) paths from the original poster to $u$ does not affect the label.

\subsection{Datasets}

We employ four Twitter datasets that consist of user follow networks, original posts, and their corresponding reposts. Each repost is classified as being propagated via either an explicit link or an implicit link, depending on whether its path can be inferred from the underlying follow network and posting timelines.

The datasets used in this study are referred to as the \textit{Higgs}~\citep{the_anatomy_of_a_scientific_rumor}, \textit{Nepal}~\citep{temporal_sequence_of_retweets_help_to_detect_influential_nodes_in_social_networks}, \textit{Turkish}~\citep{Turkish_ds}, and \textit{Ordinary}~\citep{emprical_analysis_of_the_relation_between_comunity} datasets. The Higgs dataset was collected following the announcement of the Higgs boson discovery. The Nepal dataset corresponds to Twitter activity following the 2015 Nepal earthquake. The Turkish dataset includes tweets posted in Turkish between November 2015 and January 2016. The Ordinary dataset contains a random sample of English-language reposts from 2018. For the Turkish dataset, we apply additional filtering. Whereas other datasets first collect retweets and then exhaustively compile information on the associated users, the Turkish dataset fixes the user population a priori and subsequently collects those users’ reposts. Because reposts by users outside this population are omitted, cascade graphs may be incomplete; therefore, we retain only diffusion instances for which all relevant follower relationships among the involved users are observed, ensuring a complete and accurate characterization of diffusion paths.

\begin{figure*}[tbp]
	\centering
	\includegraphics[width=1.0\textwidth]{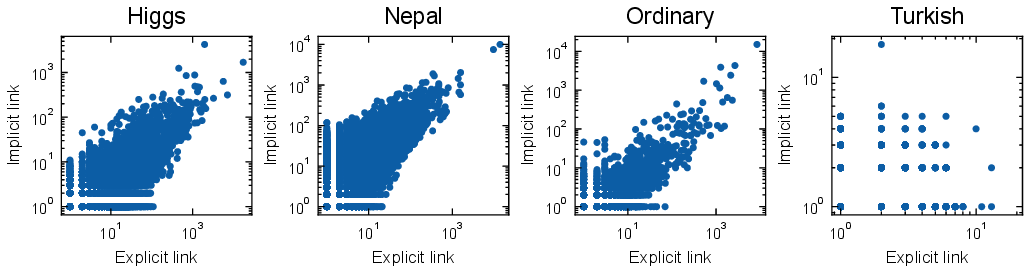}
    \caption{Correlation between implicit and explicit repost counts for each original post}
	\label{figure:implicit_explict_scatter}
\end{figure*}

Basic statistics for these datasets are provided in Table~\ref{table:datasets_summary}. The datasets range in scale from approximately 10,000 to 800,000 posts and reposts, and include the corresponding follow relationships among involved users. As shown in Table~\ref{table:datasets_summary}, between roughly 35\% and 80\% of reposts in each dataset have an explicit link, with the remainder propagating through an implicit link. Figure~\ref{figure:implicit_explict_scatter} visualizes, for each original post, the joint distribution of the number of reposts via implicit links and via explicit links. The scatter plot reveals a clear positive association between these two counts.

\begin{table}[tbp]

	\caption{Overview of the Datasets}
	\label{table:datasets_summary}
        \centering
	\begin{tabular}{ lrrrr}
    \toprule
    & Higgs & Nepal & Ordinary & Turkish \\
    \midrule
    Number of users & 456,626 & 273,213 & 111,000 & 177,611 \\
    Number of follow links & 14,855,842 & 17,818,902 & 3,130,963 & 15,940,350 \\
    Density of follow network & $7.125\times 10^{-5}$ & $2.387\times 10^{-4}$ & $2.541\times 10^{-4}$ & $4.867\times 10^{-4}$ \\
    Diameter of follow network & 17 & 15 & 17 & 9 \\
    Number of posts & 41,426 & 49,098 & 10,000 & 779,139 \\
    Number of reposts & 354,930 & 472,840 & 116,826 & 868,989 \\
    Ratio of reposts with explicit links & 0.81 & 0.34 & 0.57 & 0.38 \\
    Average cascade size & 8.57 & 9.63 & 11.68 & 1.12 \\
    
    \bottomrule
    \end{tabular}
	\end{table}
	\medskip

\section{Characterizing Implicit Link Diffusion}
\label{section:analysis_and_discussion}

\subsection{Distance-Based Analysis of Reposting Behavior}
\label{subsection:distance_analysis}

We begin our analysis by investigating how nontrivial information diffusion occurs through reposts facilitated by implicit links ({\bf RQ1}). In particular, we focus on identifying which users are more likely to engage in reposting via implicit links by analyzing their relative positions in the follow network.

In this context, the distance between a reposting user and the original post's author in the follow network serves as a proxy for the closeness of their relationship. Users with similar interests or belonging to the same communities are typically closer in the network, while those with dissimilar interests or belonging to different communities are expected to be more distant. For example, \cite{Masud_hate_topic_repost_prediction} also used the network distance from the original poster as a feature to predict repost occurrences. We note that explicit cascades are known to terminate quickly and spread only over short distances, and that external events can trigger reposts in later hops. Based on this intuition, we hypothesize that users located closer to the original source are more likely to repost through explicit links, while users at greater distances are more likely to rely on implicit links.

\begin{figure*}[tbp]
	\centering
	\includegraphics[width=1.0\textwidth]{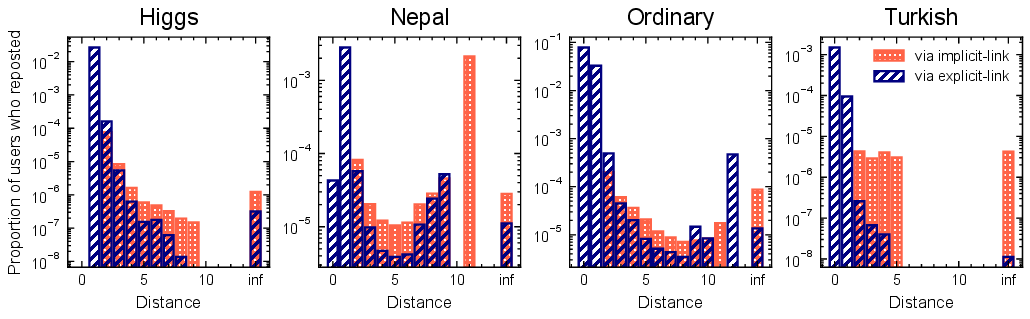}
    \caption{The proportion of reposting users among users located at a certain distance from the source}
	\label{figure:distance_2}
\end{figure*}

\begin{figure*}[tbp]
        \centering
        \includegraphics[width=1.0\textwidth]{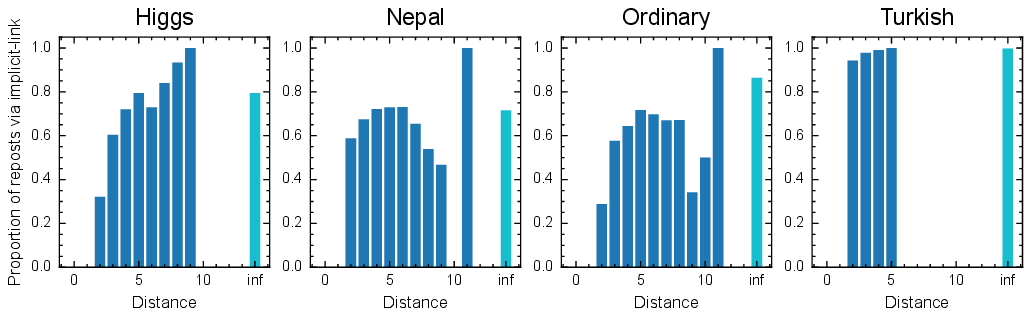}
        \caption{The proportion of reposts via implicit links by distance from the source}
	\label{figure:distance_1}
\end{figure*}

To test this hypothesis, we analyze the distribution of reposting behavior as a function of network distance from the source. Figure~\ref{figure:distance_2} shows the proportion of users who reposted via either explicit or implicit links, segmented by their distance from the original poster. Figure~\ref{figure:distance_1}, on the other hand, displays the proportion of reposts made specifically via implicit links at each distance. As expected, Figure~\ref{figure:distance_2} shows that users located closer to the source are more likely to repost overall. In contrast, Figure~\ref{figure:distance_1} reveals that among users who do repost, those at greater distances are more likely to do so via implicit links. Note that, by definition, reposts via implicit links cannot occur at distance 1.  

The observation that the proportion of implicit links varies with distance is intriguing. In general, users who are farther from the source are expected to be less interested in the information, so it is natural that the number of reposts—via both explicit and implicit links—decays with distance. However, because the decay is milder for implicit links, the share of implicit reposts grows at larger distances. For instance, when information is socially significant regardless of a user’s prior interests, people may engage in implicit reposting through search, recommendations, or trending feeds, irrespective of their distance from the source. We consider that such reposts originating from outside the follow network make up a non-negligible fraction of implicit links and help explain why the number of implicit-link reposts declines more slowly with distance. It is also possible that, due to missing observations caused by privacy settings and related factors, the probability that a repost is classified as occurring via an implicit link increases with distance. In other words, because of these data limitations, even when the true source of a repost cannot be identified, a nearby followee may have reposted independently, causing the event to be (mis)classified as an explicit link; such false explicit links are therefore more likely to occur closer to the source. We analyze differences in diffusion characteristics between implicit and explicit reposts in greater detail in later sections.

\subsection{Effects of Reposting on Future Cascade Sizes}
\label{subsection:cascade_contribution}

To clarify the effects of reposts via implicit links on the dynamics of information diffusion, we examine how such reposts contribute to the final size of diffusion cascades ({\bf RQ2}). Prior research on diffusion modeling often assumes that information propagates solely through the explicit follow network, either explicitly or implicitly. If, however, implicit link diffusion significantly contributes to cascade growth, it becomes necessary to incorporate such links into information diffusion models. Conversely, if implicit links play only a marginal role, their omission may be justified.

To evaluate the influence of individual reposts on overall cascade size, we define the \textit{Repost Contribution Index} (RCI). The RCI quantifies how much a given repost contributes to subsequent reposts and, by extension, to the growth of the entire cascade.

Let $B(r(t, v))$ denote the set of reposts made \emph{before} $r(t, v)$ by users followed by user $v$. Let $F(r(t, v))$ denote the set of reposts $r(t, w)$ such that $r(t, v) \in B(r(t, w))$, i.e., reposts that are possibly influenced by $r(t, v)$. 
Then, the RCI of repost $r(t, w)$ is recursively defined as:
\begin{equation}
\mathrm{RCI}(r(t, w)) = \sum_{r(t, v) \in F(r(t, w))} \frac{1 + \mathrm{RCI}(r(t, v))}{|B(r(t, v))|}.
\end{equation}
Note that the RCI of any repost that has never been referenced is defined to be zero, that is, $\mathrm{RCI}\bigl(r(t,v)\bigr) = 0 \quad \text{if }  \lvert F(r(t,v))\rvert = 0.$ This formulation captures the direct and indirect influence of a repost on future reposts within the cascade.

\begin{figure*}[tbp]
	\centering
	\includegraphics[width=1.0\textwidth]{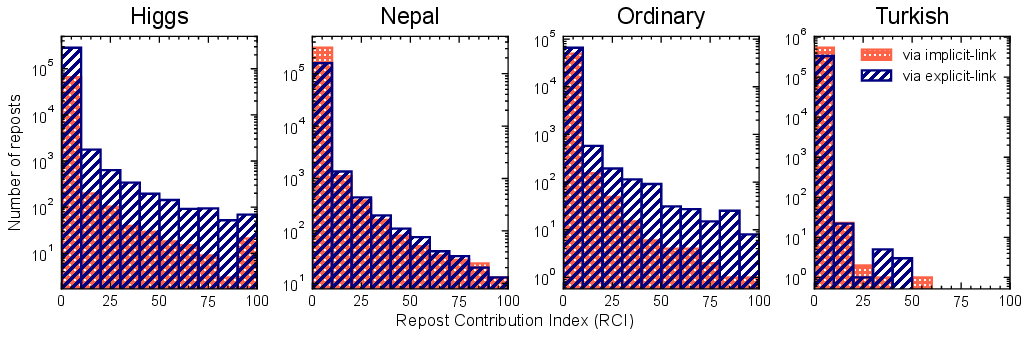}
	\caption{Distribution of the Repost Contribution Index (RCI) for reposts via explicit and implicit links}
	\label{figure:contribution}
\end{figure*}

Figure~\ref{figure:contribution} shows the distribution of RCI values for reposts via implicit and explicit links. The horizontal axis represents RCI, while the vertical axis indicates the number of reposts associated with each RCI value. As shown, the majority of reposts—regardless of whether they are via explicit or implicit links—have low RCI scores. This observation is consistent with the findings of \cite{the_structure_of_online_diffusion_networks}, who reported that most reposts have limited influence on cascade growth. Note that the presence of influential nodes, whose high centrality allows their reposts to reach broader audiences, can disproportionately amplify the overall cascade growth even when most reposts remain low in impact.

To verify that the link-type difference in RCI persists after adjusting for other metrics, we fit OLS regressions of standardized RCI on standardized covariates: follow network distance from the source, a disconnected indicator (1~=~disconnected, 0~=~connected), time since the first repost, and an explicit/implicit link indicator (1~=~explicit, 0~=~implicit). Table~\ref{table:rci_linear_regression} reports coefficients and model fit. Across all datasets except Turkish, the models explain a substantial share of variance (adjusted $R^2=0.53–0.83$), and the explicit/implicit indicator is the dominant predictor (standardized coefficients 0.67–0.96) conditional on other factors. In the Turkish dataset this pattern weakens and overall fit declines, which we attribute to its smaller mean cascade size (Table~\ref{table:datasets_summary}), leaving less headroom for individual reposts to drive downstream growth.

When comparing the two types of links, reposts via explicit links tend to exhibit higher RCI values than those via implicit links. This suggests that reposts made through explicit social connections are generally more effective in propagating information and driving cascade expansion. In contrast, while implicit link reposts occur across broader parts of the network, they appear to contribute less to the growth of diffusion cascades in terms of volume.

\begin{table}[tbp]
	\caption{Results of regression models explaining RCI, showing standardized coefficients and adjusted R² values for each dataset. Robust standard errors (HC3) in parentheses}
	\label{table:rci_linear_regression}
        \centering
\begin{tabular}{lllll}
\toprule
 & Higgs & Nepal & Ordinary & Turkish \\
\midrule
Distance from poster & -0.07 (0.0011) & 0.11 (0.0012) & 0.12 (0.0019) & -0.09 (0.0016) \\
Disconnected label & -0.01 (0.0003) & 0.13 (0.0012) & 0.10 (0.0017) & -0.05 (0.0007) \\
Repost time & 0.07 (0.0015) & -0.07 (0.0005) & -0.03 (0.0014) & 0.03 (0.0017) \\
Explicit link flag & 0.68 (0.0011) & 0.89 (0.0006) & 0.97 (0.0012) & 0.08 (0.0018) \\
Adjusted R² & 0.54 & 0.76 & 0.84 & 0.03 \\
\bottomrule
\end{tabular}
    \begin{tablenotes}\footnotesize
    \item  All dependent and independent variables were z-standardized (mean = 0, SD = 1) prior to estimation.
\end{tablenotes}
    \end{table}

\subsection{Cross-Community Diffusion via Implicit Links}
\label{subsection:community_analysis}

To investigate who receives information via implicit link diffusion, we analyze whether reposting users belong to the same community as the original poster or to a different one ({\bf RQ3}). While the previous section indicated that implicit link reposts contribute less to cascade size, they may still play an important role in broadening the audience by reaching across community boundaries. If implicit links facilitate information dissemination beyond a user's immediate social circle, they may serve as valuable pathways for spreading information to diverse groups.

For community detection, we employ the Louvain algorithm~\citep{fast_unfolding_of_communities_in_large_networks}, a widely used method in social network analysis, including studies of information diffusion~\citep{emprical_analysis_of_the_relation_between_comunity}. This algorithm partitions a network into communities characterized by a high density of links within groups and relatively fewer links between them~\citep{community_detection_in_graphs}. It is generally assumed that communities detected from follow networks reflect real-world groupings or shared interests among users.

We compared the intra-community diffusion ratio, defined as the proportion of reposts in which the reposting user and the original poster belong to the same community, between reposts via implicit links and those via explicit links.
To further understand whether the observed intra-community diffusion ratio originates from the underlying community structure—characterized by densely connected intra-community links—or from users’ behavioral tendencies to preferentially repost within their own communities, we also estimated the intra-community diffusion ratio expected under a null model. The null model generates counterfactual cascades in which community structure is absent while preserving the temporal order of reposts in the original cascade.
Starting from each original cascade graph, we constructed a counterfactual cascade as follows. For reposts via implicit links, at each step we randomly selected a user located at the same geodesic distance from the source poster as the original reposter. For reposts via explicit links, we preserved the referential mechanism by randomly choosing the reposter from among the followers of the user being explicitly referenced at that step. In this way, we produced randomized cascades that ignore community structure but retain the temporal and topological properties of the original cascades. For each original cascade, we generated ten null cascades and drew 1,000 bootstrap samples to estimate the distribution of the overall intra-community diffusion ratio. The observed standard deviation of this distribution was very small.

\begin{figure}[tbp]
	\centering
	\includegraphics[width=1.0\columnwidth]{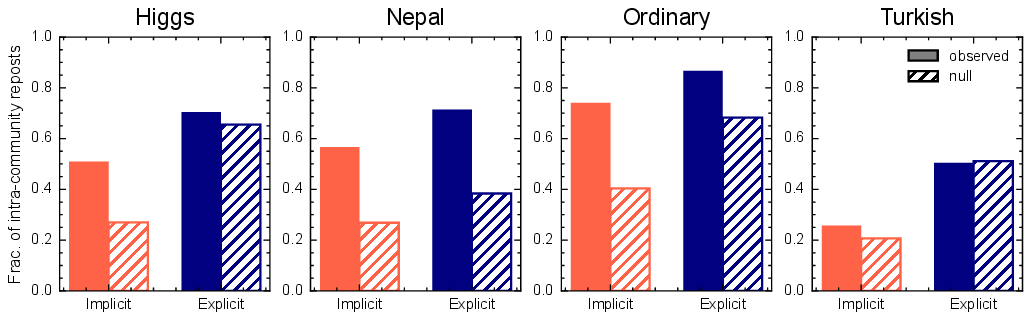}
    \caption{The fraction of reposts in which the reposting user and the original poster belong to the same community: Across all conditions, the null model’s standard deviation was very small ($< 0.01$).}
	\label{figure:community}
\end{figure}

Figure~\ref{figure:community} presents both the actual intra-community diffusion ratios and those estimated from the null model for each dataset. In all datasets, the intra-community diffusion ratio of reposts via implicit links is lower than that of explicit links. In general, intra-community diffusion tends to be high, indicating that retweets are often confined within the same community~\citep{emprical_analysis_of_the_relation_between_comunity}. Our results similarly show high intra-community diffusion ratios, confirming that reposts are likely to be trapped within communities. However, since the intra-community diffusion ratio for implicit links is lower than that for explicit links, this suggests that implicit links are more likely to carry information across diverse user groups.
Nevertheless, when compared to the null model estimates, both implicit and explicit links exhibit higher observed intra-community diffusion ratios than expected under the random baseline. This finding implies that actual user reposting behavior tends to be more confined within communities than would be predicted from network structure alone.
Overall, these results indicate that (1) reposts via implicit links are more likely to transmit information across community boundaries than those via explicit links, and (2) regardless of link type, users’ reposting behavior is generally constrained within their own communities.

This finding suggests that while implicit links may not substantially drive large-scale diffusion (as shown in Section~\ref{subsection:cascade_contribution}), they are more likely to carry information across diverse user groups. In this sense, implicit link diffusion may play an important role in mitigating echo chambers by bridging otherwise disconnected communities.

\section{User-Level Dynamics of Implicit Link Diffusion}
\label{section:user_sucepibility_and_influence}
In this section, we investigate the behavioral and structural characteristics of users who engage in or induce implicit link diffusion ({\bf RQ4}). We first analyze users as reposter, focusing on how actively they adopt content through explicit and implicit links, measured by IAR and SAR, respectively. We then examine users as post authors, analyzing how their posts are retweeted by others using a newly introduced metric, Repost via Explicit-link Rate (RER), which captures the extent to which their posts are reposted through implicit links. These user-level analyses provide insights into how individual user behaviors give rise to cascades that include implicit links.

\subsection{Susceptibility to Implicit Links}
\label{section:user_sucepitibility}
Next, we examine the characteristics of users who are more likely to participate in information diffusion via implicit links ({\bf RQ4}). For this analysis, we adopt two metrics introduced by \cite{luceri202IAR_SAR}: the IAR and the SAR. In their framework, IAR measures a user's susceptibility to adopting information via explicit links—that is, from posts shared by their social connections. In contrast, SAR measures the user's susceptibility to adopting information via implicit links, which are not mediated by direct social relationships.

We apply these metrics to our datasets to analyze the distribution of IAR and SAR across users. Formally, let $T^{\text{adopted}}_u$ denote the set of tweets reposted by user $u$, and let $T^{\text{exposed}}_u$ denote the set of tweets to which user $u$ was exposed via explicit links. Then, IAR and SAR are defined as:

\begin{equation}
\mathrm{IAR}(u) = p(adopted \mid exposed) = \frac{|T^{\text{adopted}}_u \cap T^{\text{exposed}}_u|}{|T^{\text{exposed}}_u|}
\end{equation}

\begin{equation}
\mathrm{SAR}(u) = 1 - p(exposed \mid adopted) = 1 - \frac{|T^{\text{exposed}}_u \cap T^{\text{adopted}}_u|}{|T^{\text{adopted}}_u|}.
\end{equation}

\begin{table}[tbp]
	\caption{The number of users used for the analyses of IAR and SAR}
	\label{table:iar_sar_filter}
        \centering
    	\begin{tabular}{ lrrrr}
        \toprule
        & Higgs & Nepal & Ordinary & Turkish \\
        \midrule
        Num. of reposted users (original) & 228,556 & 257,268 & 112,199 & 63,729 \\
        Num. of users who reposted at least five times & 7,779 & 13,585 & 49 & 20,024 \\
        Num. of users who have at least one mutual follow link & 3,903 & 8,520 & 22 & 17,749 \\
        \bottomrule
        \end{tabular}
        \end{table}

\begin{figure*}[tbp]
\centering
    \subfigure[IAR Distribution] {
        \centering
        \includegraphics[width=1.0\textwidth]{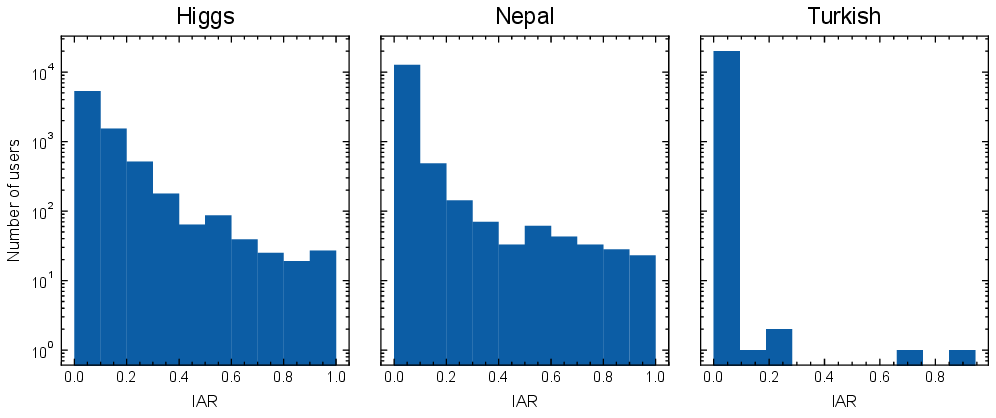}
	\label{figure:iar_distribution}
    }
    \subfigure[SAR Distribution] {
        \centering
        \includegraphics[width=1.0\textwidth]{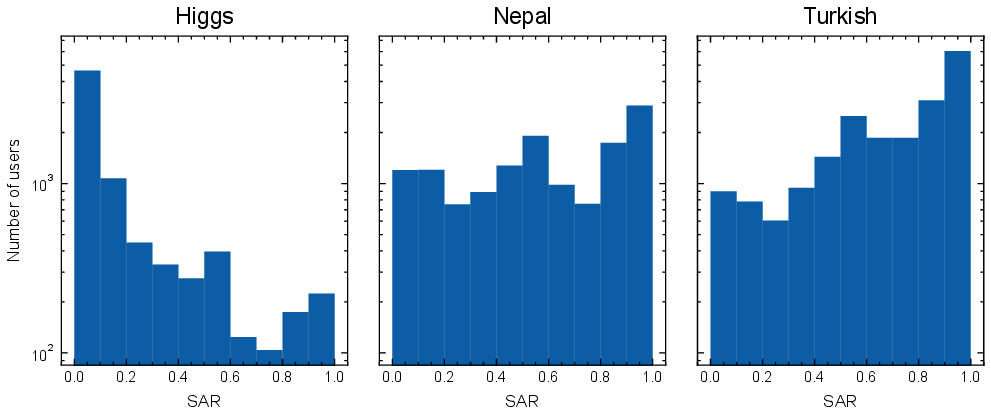}
	\label{figure:sar_distribution}
    }
        \caption{Distribution of IAR and SAR across users in the Higgs, Nepal, and Turkish datasets}
        \label{figure:iar_sar_distribution}
\end{figure*}

Figure~\ref{figure:iar_sar_distribution} shows the distributions of IAR and SAR in the Higgs, Nepal, and Turkish datasets. To ensure statistical reliability, we exclude users who reposted fewer than five times. The numbers of users after the filtering process are summarized in Table~\ref{table:iar_sar_filter}.
Note that we omit the Ordinary dataset from this analysis, as it contains too few users meeting the threshold to reliably compute IAR and SAR.

The results reveal considerable variation in users’ SAR values, suggesting that susceptibility to implicit link diffusion is not random, but shaped by user-specific characteristics.
The differences in trends observed across the individual datasets are likely attributable to contexts of data-collection (academic news, non-event, and disaster) and differences in the data‑collection strategies (keyword‑based sampling (Higgs, Nepal) versus user‑based sampling (Turkish)).

\begin{figure*}[tbp]
        \centering
        \includegraphics[width=1.0\textwidth]{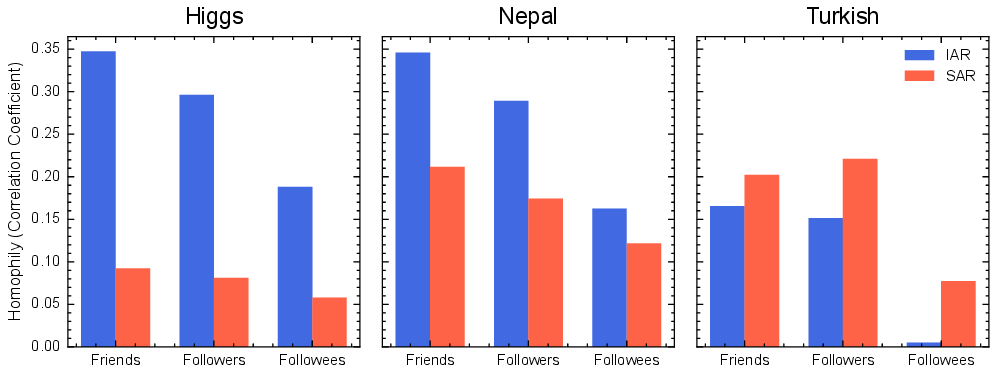}
        \caption{Homophily of IAR and SAR. Spearman correlation between a user's IAR/SAR and the average IAR/SAR of their neighbors under different types of adjacency (followers, followees, and mutual connections). All correlations were significant ( $p<0.01$ ), except for the correlation of Turkish SAR with followers}
	\label{figure:iar_sar_homopholy}
\end{figure*}

To further investigate this, we assess homophily in user susceptibility—that is, whether users with high IAR or SAR tend to be socially proximate to others with similar values. Figure~\ref{figure:iar_sar_homopholy} reports the Spearman rank correlation between a user's IAR (or SAR) and the average IAR (or SAR) of their neighbors. We consider three types of user adjacency: (1) users followed by a given user (followees), (2) users who follow the user (followers), and (3) users with mutual follow relationships (friends). 

As shown in Figure~\ref{figure:iar_sar_homopholy}, we observe positive, and weak/moderate, but statistically significant correlations in all adjacency types for both IAR and SAR, indicating moderate-level of homophily in user susceptibility. These findings are consistent with those reported by \cite{luceri202IAR_SAR}, and support the idea that susceptibility to both influence-driven and spontaneous adoption is a socially embedded trait.

Across all datasets and adjacency types, the correlation is strongest among mutual connections (friends), reflecting that stronger social ties are associated with higher behavioral similarity. In comparison, \cite{luceri202IAR_SAR} report correlations in the 0.30–0.58 range, whereas the effects we observe are somewhat weaker. One possible explanation is that while \cite{luceri202IAR_SAR} inferred user relationships from mention interactions—often representing stronger social bonds—we use follow relationships, which may include weaker or inactive ties. This aligns with the interpretation that homophily is more pronounced between users connected via stronger relational signals. Note that public figures tend to behave differently from ordinary users on social media, forming distinct types of social connections and engaging in different patterns of interaction. Therefore, a detailed examination of differences by type of social connection may need to control for the presence of public figures.
Another consistent observation is that IAR exhibits stronger homophily than SAR, which is also in line with the findings of \cite{luceri202IAR_SAR}.

\begin{figure*}[tbp]
\centering
\subfigure[IAR distribution in network] {
    \includegraphics[width=1.0\textwidth]{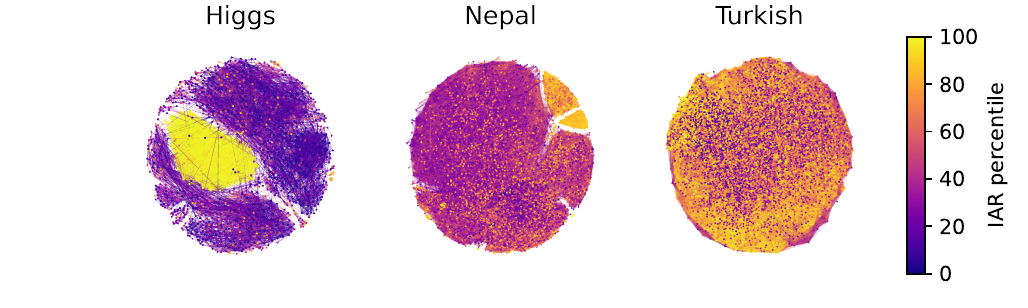}
    \label{figure:iar_network}
  }
\subfigure[SAR distribution in network] {
        \includegraphics[width=1.0\textwidth]{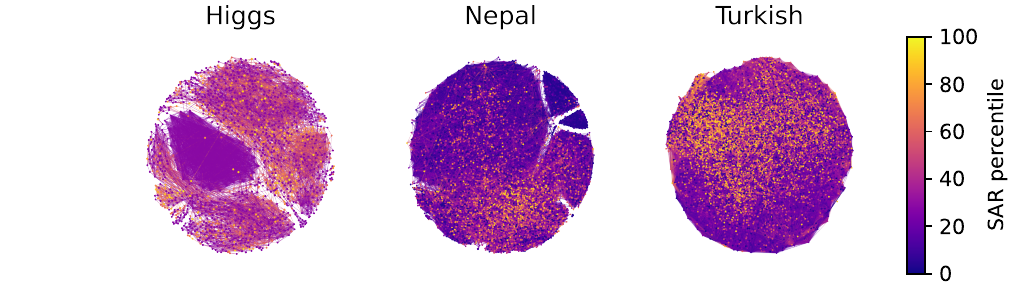}
  \label{figure:sar_network}
}
\caption{ IAR and SAR distribution in network: Nodes represent users (who reposted at least five times). Links represent mutual-follower relationships, and node colors indicate the user’s IAR (SAR) percentile. The network has been decomposed into its k-core (k=3) and arranged using the ForceAtlas2 force-directed layout}
\label{figure:iar_sar_network}
\end{figure*}

We next investigate the global distribution of IAR and SAR in network structures. In Figure~\ref{figure:iar_sar_network}, we show the IAR and SAR distributions in the mutual-follower network. In the figures, we are able to observe structural biases. There are groups of higher IAR (or SAR) users and lower IAR (or SAR) users. Notably, IAR exhibits more pronounced clustering than SAR, with clearer community-level concentrations of high and low values. In particular, for the Higgs and Nepal datasets, users within the same community tend to share similar values. In the Turkish dataset, by contrast, two groups of users—those with higher and lower values—appear to be connected through a gradual gradient rather than a distinct separation. This pattern accords with our homophily results. The previous analysis showed first-order similarity between directly connected users. However, there may exist a similarity between users in second-order or higher-order relationships.

\begin{figure*}[tbp]
\centering
\subfigure[IAR distance correlation]{
        \includegraphics[width=1.0\textwidth]{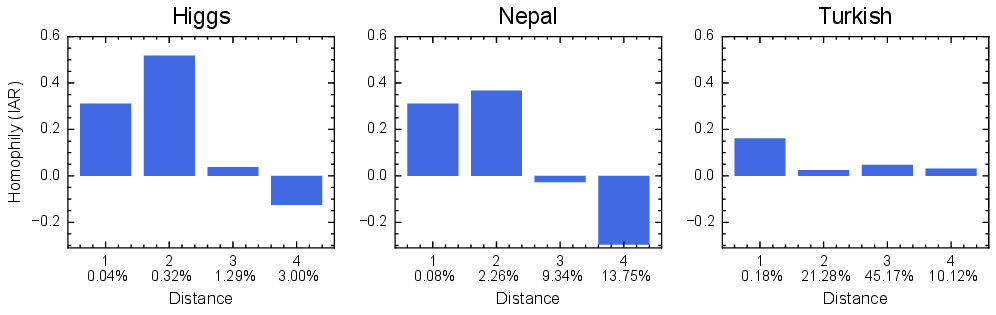}
}
\subfigure[SAR distance correlation] {
        \includegraphics[width=1.0\textwidth]{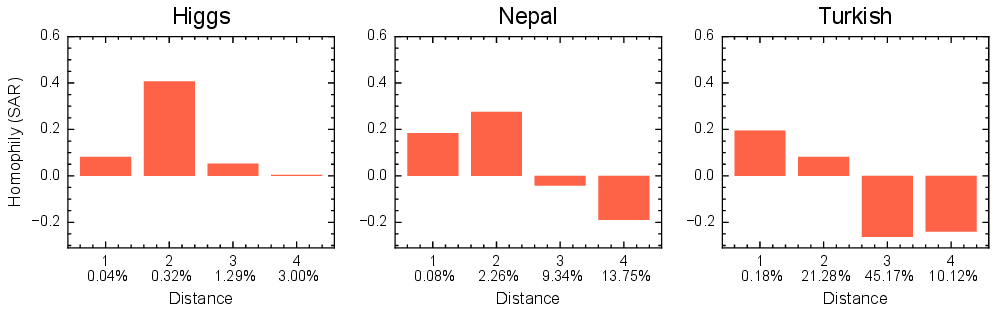}
}
        \caption{Distance correlation of IAR and SAR: For each dataset (Higgs, Nepal, Turkish), bars show the Spearman rank correlation between a user’s IAR/SAR and the average IAR/SAR of the users who are exactly $d$ hops away from that user in the mutual‑follower network (with $d=1, 2, 3, 4$). The lower row of x-labels represents the mean percentage of users present at each distance}
        \label{figure:distance_correlation_of_iar_sar}
\end{figure*}

To validate higher-order similarity of IAR and SAR, we calculated homophily extended up to a distance of four. We calculated Spearman's rank correlation between users' IAR (or SAR) and the average IAR (or SAR) of users at exactly specific distances in mutual-follower networks. In Figure~\ref{figure:distance_correlation_of_iar_sar}, we display these correlations for each exact distance together with the proportion (share) of users located at that distance. As shown in Figure~\ref{figure:distance_correlation_of_iar_sar}, there are slightly higher correlations with two-hop users than one-hop users in the Higgs and Nepal datasets. 
This elevated two-hop correlation suggests the presence of monophily~\citep{Altenburger_2018}, individuals with extreme attribute preferences unrelated to their own attributes.
The absence of observed moderate monophily in the Turkish dataset may be attributable to the fact that the Turkish dataset represents a densely connected network, in which as many as 21\% of users are reachable within a distance of two, compared with 0.3\% in Higgs and 2.2\% in Nepal.

Taken together, our findings demonstrate that \cite{luceri202IAR_SAR}'s results are robust across different datasets and definitions of social connections. Moreover, we find that the degree of homophily in IAR and SAR increases with the strength of the connection between users. This suggests that incorporating tie strength into neighborhood-based modeling may improve the accuracy of predicting user susceptibility to different types of information diffusion. In addition to confirming Luceri et al.'s observation, our analysis reveals strong monophily in both IAR and SAR, suggesting the importance of incorporating higher-order neighbors—beyond just 1-hop connections—such as by employing graph convolutional neural networks~\citep{Kipf_gcn}, which could play a critical role in enhancing predictive accuracy.

\subsection{Induction of Implicit Link Diffusion}
\label{section:user_influence}

Finally, we examine the characteristics of users who are more likely to \textit{induce} implicit link diffusion—that is, users whose posts are frequently reposted by others who have no explicit link.
While IAR and SAR quantify \emph{consumer-side} susceptibility (how likely a user adopts), we introduce a complementary \emph{producer-side} metric that captures how a user's posts \emph{induce} diffusion via explicit or implicit links.
Identifying such users is valuable for understanding how information propagates beyond social boundaries, and could inform strategies for viral marketing or outreach that leverage non-social pathways.

We introduce the Repost via Explicit link Rate (RER), which measures the proportion of a user's received reposts that originate from explicit links. A lower RER indicates that a larger share of reposts comes from users not socially connected to the original poster, suggesting a stronger tendency to induce implicit link diffusion. 
RER of user $u$ is defined as follows:
\begin{align}  
\mathrm{RER}(u) &\notag\\
&= p(\text{reposted via explicit link} \mid reposted)\notag\\
&= \frac{\left|\left\{r(t,v)\;:\; a(t)=u,\ \exists x\ \text{s.t.}\ (x,v)\in E_t\right\}\right|} {\left|\left\{r(t,v)\;:\; a(t)=u\right\}\right|}.
\end{align}

\begin{table}[tbp]
	\caption{The number of users used for the analysis of RER}
	\label{table:RER_filter}
        \centering
        \begin{tabular}{lrrrr}
        \toprule
        & Higgs & Nepal & Ordinary & Turkish \\
        \midrule
        Num. of users whose posts were reposted (original) & 41,426 & 18,983 & 9,900 & 34,503 \\
        Num. of users who were reposted at least five times & 7,012 & 8,971 & 1,413 & 15,537 \\
        Num. of users who have at least one mutual follow link & 5,732 & 4,979 & 571 & 14,535 \\
        \bottomrule
        \end{tabular}
        \end{table}
        \medskip

\begin{figure*}[tbp]
        \centering
        \includegraphics[width=1.0\textwidth]{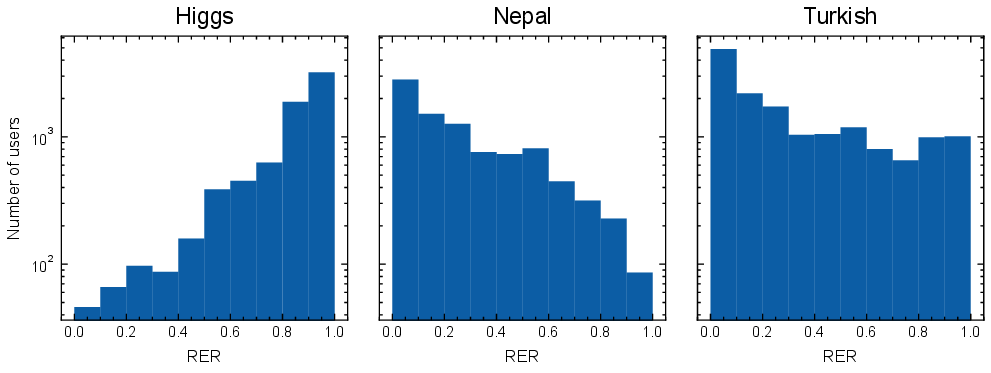}
        \caption{Distribution of Repost via Explicit link Rate (RER) across users. Lower values indicate stronger tendencies to induce implicit link diffusion}
	\label{figure:RER_distribution}
\end{figure*}

Figure~\ref{figure:RER_distribution} shows the distribution of RER values across users in the Higgs, Nepal, and Turkish datasets. To ensure statistical reliability, we exclude users who have been reposted fewer than five times. The results of this filtering are summarized in Table~\ref{table:RER_filter}.
The distribution of RER exhibits different patterns depending on the dataset. These differences are likely due to the contexts of data collection conditions. In all datasets, some users have very low RER values. This suggests the existence of a subset of users who are particularly effective at triggering reposts from users outside their social network.

\begin{figure*}[tbp]
        \centering
        \includegraphics[width=1.0\textwidth]{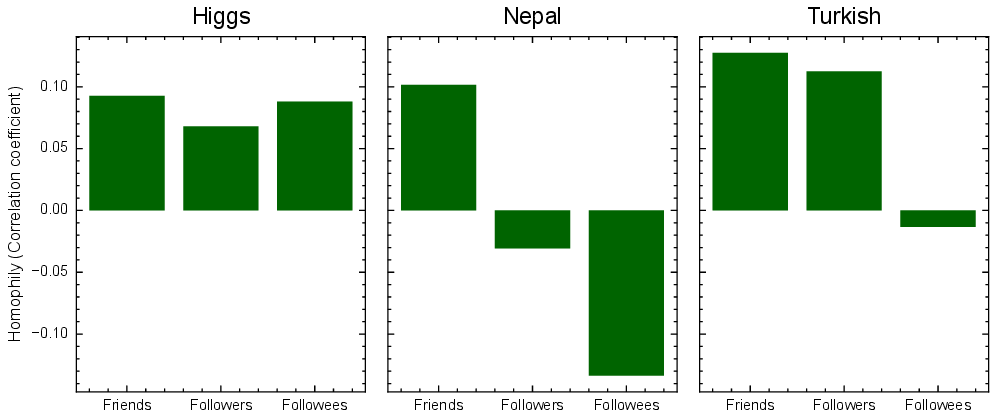}
        \caption{Homophily of RER. Spearman correlation between a user's RER and the average RER of their neighbors across different types of adjacency.  All correlations are significant ( $p<0.01$ ), except for the correlation of Nepal Followers and Turkish Followees}
	\label{figure:RER_homopholy}
\end{figure*}

We further examine whether users with similar RER values tend to be socially proximate—that is, whether the ability to induce implicit link diffusion exhibits homophily. Following the same procedure as in Section~\ref{section:user_sucepitibility}, we compute the Spearman rank correlation between a user’s RER and the average RER of their neighbors under three types of adjacency: (1) followees, (2) followers, and (3) mutual followers (friends). Figure~\ref{figure:RER_homopholy} presents the results.  In the case of mutual connection (friends), the correlation was statistically significant, but only a very weak association was observed. Compared with IAR/SAR, homophily is much weaker for RER; therefore, this result suggests that if the goal is to characterize post authors or to find influencers who are adept at implicit diffusion, information about directly adjacent nodes may be of limited use.

\begin{figure*}[tbp]
        \centering
        \includegraphics[width=1.0\textwidth]{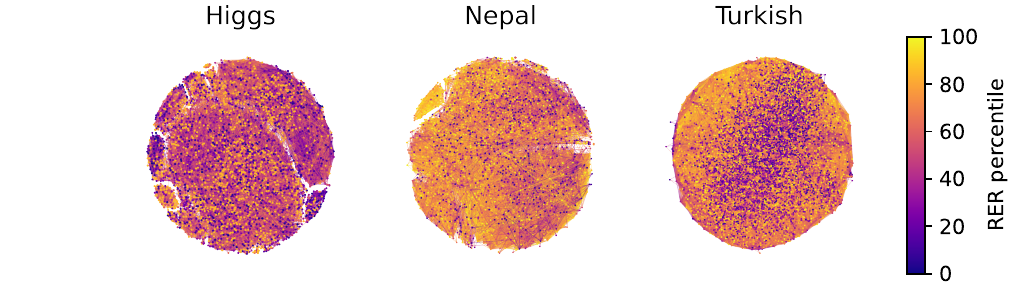}
        \caption{RER distribution in network: Nodes represent users (who have been reposted at least five times). Links represent mutual-follower relationships, and node colors indicate the user’s RER percentile. The network has been decomposed into its 3-core (k=3) and arranged using the ForceAtlas2 force-directed layout }
	\label{figure:rer_graph_vis}
\end{figure*}

We also examined the global distribution of RER.
Figure~\ref{figure:rer_graph_vis} shows the RER distribution in the mutual-follower network. Although weaker than the IAR pattern in Figure~\ref{figure:iar_sar_distribution}, RER likewise exhibits network-level bias. In the Higgs dataset, communities tend to display similar RER values, indicating community-level clustering. In the Turkish dataset, by contrast, the network appears to comprise two groups—one with generally higher RER and one with lower RER—connected by a gradual gradient rather than a sharp boundary.

\begin{figure*}[tbp]
        \centering
        \includegraphics[width=1.0\textwidth]{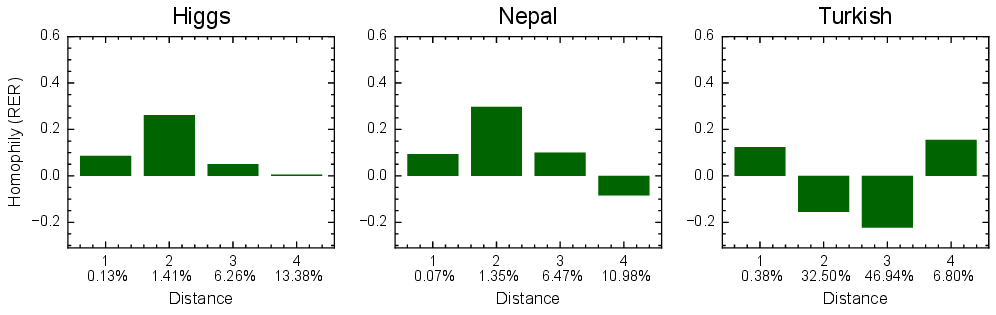}
        \caption{Distance correlation of RER: For each dataset (Higgs, Nepal, Turkish), bars show the Spearman rank correlation between a user’s RER and the average RER of the users who are exactly $d$ hops away from that user in the mutual‑followee network (with $d=1, 2, 3, 4$). The lower row of x-labels represents the mean percentage of users present at each distance }
	\label{figure:rer_distance_correlation}
\end{figure*}

Subsequently, we examined monophily, as it may give rise to bias within the network. In Figure~\ref{figure:rer_distance_correlation}, we show the correlation of users' RER and the average RER of users who are located at certain distances. As shown in the figure, the result of RER distance correlation shows a trend similar to IAR and SAR.
There are peaks in two-hop users, which implies the existence of monophily.
In the Turkish dataset, we observed a different pattern. It might be caused by a dense network in the Turkish dataset. Approximately 80\% of users in the dataset are located within 2-hop or 3-hop distances. This suggests that the overall structure of the network may influence the distance correlation.

These results also imply that the RER of a user may be inferred from the behavior of their social neighbors, especially higher-order neighbors. This observation may be useful for tasks such as identifying potential implicit link influencers or modeling the spread of information in hybrid diffusion networks that incorporate both social and non-social pathways.

\section{Discussion}
\label{sec:discuss}

\subsection{Implications}

Our findings reveal that a significant portion of information diffusion on social media occurs through implicit links, rather than through direct social ties. In particular, users who are located farther away from the original poster in the follow network are more likely to repost content via implicit links (Section~\ref{subsection:distance_analysis}). This implies that traditional diffusion models that assume propagation only through explicit links—such as the Independent Cascade Model or Linear Threshold Model~\citep{Kempe03:Maximizing}—may fail to accurately reproduce the reach and structure of real-world diffusion cascades. Future diffusion models should incorporate mechanisms that reflect the presence and characteristics of implicit link diffusion, especially when modeling long-range information spread.

While reposts via implicit links tend to have a lower contribution to cascade growth compared to reposts via explicit links (Section~\ref{subsection:cascade_contribution}), they serve a crucial structural role in bridging otherwise disconnected communities. Our analysis shows that implicit links are more likely than explicit links to connect users across community boundaries. Even so, relative to the null model the share of cross-community diffusion remains low, indicating that implicit diffusion—like explicit diffusion—exhibits high locality within communities (Section~\ref{subsection:community_analysis}). This finding offers important implications for mitigating echo chambers~\citep{cinelli2021echo}, which often arise when information is confined within densely connected communities. Enhancing diffusion through implicit links—such as through algorithmic recommendations or content trending mechanisms—could help promote broader exposure and reduce ideological polarization.

At the user level, we observe systematic variation in susceptibility to implicit link diffusion (Section~\ref{section:user_sucepitibility}) and in the ability to induce it (Section~\ref{section:user_influence}). Users with higher SAR or lower RER are more involved in nontrivial diffusion, and these characteristics exhibit modest but statistically significant homophily and  monophily. This suggests that both susceptibility and inducement are not purely individual traits but are socially embedded. 
These findings point to the potential utility of network-aware prediction models that incorporate local social structure and tie strength to identify users who are more likely to participate in or trigger diffusion via implicit links. However, because the observed correlations are weak, the potential for these signals to deliver meaningful predictive gains requires careful empirical validation.

While monophily was observed across all metrics, RER exhibited lower homophily compared to IAR and SAR. This difference may be explained by the nature of the metrics themselves: IAR and SAR reflect the characteristics as an information consumer (i.e., how individuals interact with information), whereas RER captures the characteristics as an information producer (i.e., the kinds of posts they author).
Individuals with similar traits as information producers, as measured by RER, may not necessarily form direct connections with each other; instead, they may be more likely to share common audiences or fans. Consequently, homophily may tend to be lower in RER, whereas monophily could be relatively high. In contrast, IAR and SAR, which reflect information consumer characteristics, suggest that individuals with similar consumption patterns may be more likely to connect directly, possibly resulting in both higher homophily and monophily. These findings may highlight the different network structures associated with information production and consumption behaviors.

Overall, our results highlight the roles of implicit links in information diffusion, roles that are closely related to the structure of the network. Although the scale of information diffusion facilitated by implicit links may not be large, these links have the distinctive ability to bridge across communities. Moreover, users who contribute to the formation of implicit links tend to be unevenly distributed within the network.

\subsection{Limitations and Future Directions}

While our study provides new insights into the dynamics and user-level properties of implicit link diffusion, it has several limitations that suggest avenues for future research.

First, although we identified structural and behavioral patterns associated with implicit link diffusion, we did not explicitly examine the underlying mechanisms through which users encounter content outside their social networks. Such mechanisms may include algorithmic recommendations, search queries, hashtags, or external sources such as websites and news aggregators. Future work incorporating richer interaction data—such as likes, views, impressions, or browsing histories—would help clarify how implicit links are formed.

Second, our analysis is descriptive and does not propose or evaluate new diffusion models. Incorporating implicit link diffusion into formal information diffusion models remains an open research question. Extending conventional models to include non-social exposure pathways, and simulating their impact on diffusion dynamics and intervention strategies (e.g., seeding, link recommendation, or content prioritization), is a promising direction for future work.

Third, while we analyzed four datasets covering different social events and geography, the generalizability of our findings to other platforms and cultures remains to be tested. In particular, the influence of platform-specific algorithms (e.g., recommendation engines on TikTok or YouTube) on implicit link diffusion may differ significantly from the dynamics observed on Twitter.

Lastly, our definition of explicit relationships is based solely on the follow network. While this is a reasonable and widely used approximation, other relational signals such as mentions, replies, or co-engagement patterns may offer complementary perspectives. Exploring multiple types of social ties and their relative strength could further refine our understanding of how information flows in hybrid social systems.

\section{Conclusion}
\label{section:conclusion}
In this study, we investigated the dynamics of nontrivial information diffusion via implicit links on social media platforms. Using four Twitter datasets across different social events and geography, we analyzed how reposts that occur outside explicit follow relationships contribute to the overall spread of information.

Our findings reveal the multifaceted nature of implicit link diffusion. We showed that reposts via implicit links are more likely to originate from users who are socially distant from the original source (RQ1). These reposts contribute less to the overall growth of information cascades than those via explicit links (RQ2). Both implicit and explicit links tend to remain within communities; however, compared to explicit links, implicit links play a structurally important role by enabling information to traverse community boundaries (RQ3). This cross-community diffusion may help reduce echo chambers and broaden audience reach.

At the user level, we found substantial variation in susceptibility to implicit link diffusion. Using the SAR and IAR, we demonstrated that reposting behavior via implicit or explicit links is not random but socially patterned. Users tend to be connected to others with similar susceptibility profiles (homophily), while individuals with extreme, potentially attribute-independent preferences are also present (monophily) (RQ4). Furthermore, by introducing the RER, we identified users who are more likely to induce implicit link diffusion. While this trait exhibits weaker homophily than susceptibility, it shows strong monophily, indicating that the capacity to trigger reposts from outside one’s network is partially predictable from social surroundings.

Taken together, these findings offer new insights into the limitations of conventional information diffusion models that assume propagation only via explicit links. Our study underscores the need for models that account for nontrivial diffusion paths and for systems that leverage implicit link diffusion to enhance both the reach and diversity of information spread.

Looking ahead, future work should focus on identifying the external mechanisms—such as algorithmic recommendations or off-platform exposure—that generate implicit links, and on building predictive models that incorporate both social and non-social pathways of diffusion. Such efforts will enable more accurate simulations of information spread and facilitate the design of more effective interventions for applications including misinformation containment, viral marketing, and public awareness campaigns.

\section*{Statements and Declarations}
\backmatter
\bmhead{Funding} This work was partly supported by JSPS KAKENHI JP25K03105, JP23K28069, and JP22K11990.

\bmhead{Competing Interests}
The authors have no competing interests to declare that are relevant to the content of this article.

\bmhead{Data Availability}

The Higgs, Nepal, and Turkish datasets are publicly available (\url{https://snap.stanford.edu/data/higgs-twitter.html}; 
\url{https://doi.org/10.5281/zenodo.2587475}; \url{https://github.com/zezealp/twitter-dataset}). The Ordinary dataset is not publicly available due to Twitter’s policy.

\bmhead{Author contributions}
Conceptualization: Y.T, S.T, K.W;
Data curation: Y.T;
Formal analysis: Y.T;
Investigation: Y.T;
Methodology: Y.T, S.T;
Project administration: S.T;
Resources: S.T;
Software: Y.T;
Supervision: S.T, K.W;
Validation: Y.T, S.T;
Visualization: Y.T;
Writing – original draft: Y.T;
Writing – review \& editing: Y.T, S.T, K.W;
Funding acquisition: S.T, K.W;





\bibliography{sn-bibliography}

\end{document}